\documentclass[letterpaper,english,reprint,superscriptaddress,aps]{revtex4-1}
\usepackage[T1]{fontenc}
\usepackage[latin9]{inputenc}
\usepackage{color}
\usepackage{babel}
\usepackage{units}
\usepackage{amsmath}
\usepackage{amssymb}
\usepackage{graphicx}
\usepackage[unicode=true,pdfusetitle,
 bookmarks=true,bookmarksnumbered=false,bookmarksopen=false,
 breaklinks=false,pdfborder={0 0 1},backref=false,colorlinks=true]
 {hyperref}
\hypersetup{
 citecolor=blue, linkcolor=blue, urlcolor=blue}

\newcommand{\be}{\begin{equation}}
\newcommand{\ee}{\end{equation}}
\newcommand{\bea}{\begin{eqnarray}}
\newcommand{\eea}{\end{eqnarray}}

\newcommand{\fig}[1]{Fig.~\ref{#1}}

\newcommand{\e}{\varepsilon}
\newcommand{\w}{\omega}
\newcommand{\s}{\sigma}
\newcommand{\up}{\uparrow}
\newcommand{\down}{\downarrow}

\makeatletter

\pdfpageheight\paperheight
\pdfpagewidth\paperwidth

\makeatother

\begin{document}

\title{Spectral properties and the Kondo effect of cobalt adatoms on silicene}

\author{I. Weymann}
\email{weymann@amu.edu.pl}
\affiliation{Faculty of Physics, Adam Mickiewicz University, 61-614 Pozna\'{n}, Poland }

\author{M. Zwierzycki}
\email{maciej.zwierzycki@ifmpan.poznan.pl}

\selectlanguage{english}%

\author{S. Krompiewski}

\affiliation{Institute of Molecular Physics, Polish Academy of Sciences, 60-179
Pozna\'{n}, Poland}

\date{\today}
\begin{abstract}
In terms of the state-of-the-art first principle computational methods
combined with the numerical renormalization group technique the spectroscopic properties
of Co adatoms deposited on silicene are analyzed.
By establishing an effective low-energy Hamiltonian based on
first principle calculations, we study the behavior of the local density of states
of Co adatom on external parameters, such as magnetic field and gating.
It is shown that the Kondo resonance with the Kondo temperature
of the order of a few Kelvins can emerge
by fine-tuning the chemical potential.
The evolution and splitting of the Kondo peak with external magnetic field is also analyzed.
Furthermore, it is shown that the spin polarization of adatom's spectral function
in the presence of magnetic field can be relatively large,
and it is possible to tune the polarization and its sign by electrical means.
\end{abstract}

\maketitle

\section{Introduction}

The discovery of graphene, a two-dimensional (2D) honeycomb lattice
of carbon atoms, in 2004 \citep{Novoselov:sc04} spurred an interest
in other 2D materials, especially those sharing graphene's crystal
structure. The search was motivated by the hope that such materials
would also share the defining feature of graphene, \emph{i.e.} the
presence of Dirac cones in their electronic structures \citep{neto:rmp08}.
The suppression of backscattering characteristic of Dirac fermions
should then lead to similarly high carrier mobilities \citep{Novoselov:nat05,Zhang:nat05,Bolotin:prl2008,Du:NNano2008}
and possible use of graphene siblings in ultra-fast electronic devices
\citep{Lin:NanoLet2009,Liao:Nature2010,Lin:Science2011,Schwierz:NatNano2010,Kim:nanotech2012}.
The continuously growing list of elements and compounds for which
the existence of such graphene analogues was predicted theoretically
and in some cases confirmed experimentally includes: Si (so called
\emph{silicene}) \citep{Cahangirov:prl2009,Lebegue:prb2009,Vogt:prl2012,Feng:NanoLett2012,
Lin:AppPhysExpr2012,Jamgotchian:jp2012,Chiappe:AdvMat2012},
Ge (\emph{germanene}) \citep{Cahangirov:prl2009,Lebegue:prb2009},
Sn (\emph{stanene}) \citep{Garcia:JPhysChemC2011,Saxena:SciRep2016},
Al (\emph{aluminene})\emph{ }\citep{Kamal:NJPhys2015,Yuan:ApplSurfSci2017}
and hexagonal BN (\emph{white graphene}) \citep{Nag:ACSano2010}.

Of these, silicene is particularly interesting thanks to its compatibility
with the current Si-based electronic technology. The first proof of
the concept field effect transistor made out of silicene has already been
demonstrated \citep{Tao:NatNano2015}. Many of the characteristic
properties of graphene are predicted to be present also in its silicon counterpart.
The band structure of the free-standing silicene
exhibits the expected Dirac cones \citep{Cahangirov:prl2009}, which
can be preserved also on suitably selected substrates \citep{Quhe:SciRep2014}.
The zigzag edges of silicene nanoribbons are predicted to be spin-polarized
\citep{Cahangirov:prl2009,Weymann:prb2015,Weymann:prb2016} just like
in the case of graphene \citep{neto:rmp08}. However, some notable
differences also exist between the two materials. While graphene is
planar, in silicene the two sublattices are shifted vertically (\emph{buckling})
because of larger in-plane lattice constant (weaker $\pi$ bonds)
and the element's preference for forming $sp^{3}$ hybrids (no graphite
analogue exists for Si). Consequently the $\pi$ and $\sigma$ bands
are hybridized. The spin-orbit (SO) interaction in silicene is three
orders of magnitude stronger than in graphene and is responsible for
a small $\unit[1.5]{meV}$ gap in the electronic structure \citep{Liu:prl2011,PhysRevB.84.195430}.
This may lead to the realization of the spin Hall effect in experimentally
accessible temperatures. The symmetry breaking effect due to buckling
opens an interesting possibility of fine tuning the gap using vertical
electric field \citep{Drummond:prb2012,Ni:NanoLett2012}.

Another way of affecting the properties of 2D materials
is by deposition of magnetic adatoms on the surface.
In fact, the presence of impurities
has been invoked to explain the
spin relaxation time in graphene \cite{Kochan2014Mar}.
Besides modification of material properties,
individual magnetic adatoms themselves can pose very interesting
objects to study. This is because
strong coupling between localized states
of adatoms and the band of 2D material can result in
various nontrivial effects. One of such effects,
which has been widely studied in the context of quantum dots and molecules,
is undoubtedly the Kondo effect \cite{Kondo_Prog.Theor.Phys32/1964}.
In this effect the magnetic moment of confined electrons,
either in an adatom or a quantum dot,
becomes screened by surrounding mobile electrons.
This results in the formation of a resonance
in the local density of states at the Fermi level \cite{Hewson_book}.
The Kondo effect due to the presence of magnetic adatoms
has already been considered in the case of graphene \cite{Wehling2010Mar,Fritz2013Feb}.
Moreover, the spectroscopic properties of Co adatoms
on graphene have also been examined \cite{Brar2011Jan},
and recently the presence of the Kondo effect has been reported \cite{Ren2014Jun}.
Thus, while there are several considerations of spectroscopic properties
of magnetic adatoms on graphene, not much is known about the
spectral features and, in particular, the Kondo effect for other 2D materials,
such as silicene. The aim of this paper is therefore to shed light
on the physics of Co adatoms on silicene, with an emphasis on the Kondo regime.

The paper is organized as follows. In Sec. II we discuss
the first principles methods used to determine the
lowest energy geometry
and the density of states (DOS) of silicene with Co adatom.
Section III is devoted to numerical renormalization group calculations.
First, we formulate the effective Hamiltonian,
describe the method and then discuss the numerical results.
The paper is summarized in Sec. IV.

\section{First principles calculations}

\begin{figure}[t]
\includegraphics[width=0.8\columnwidth]{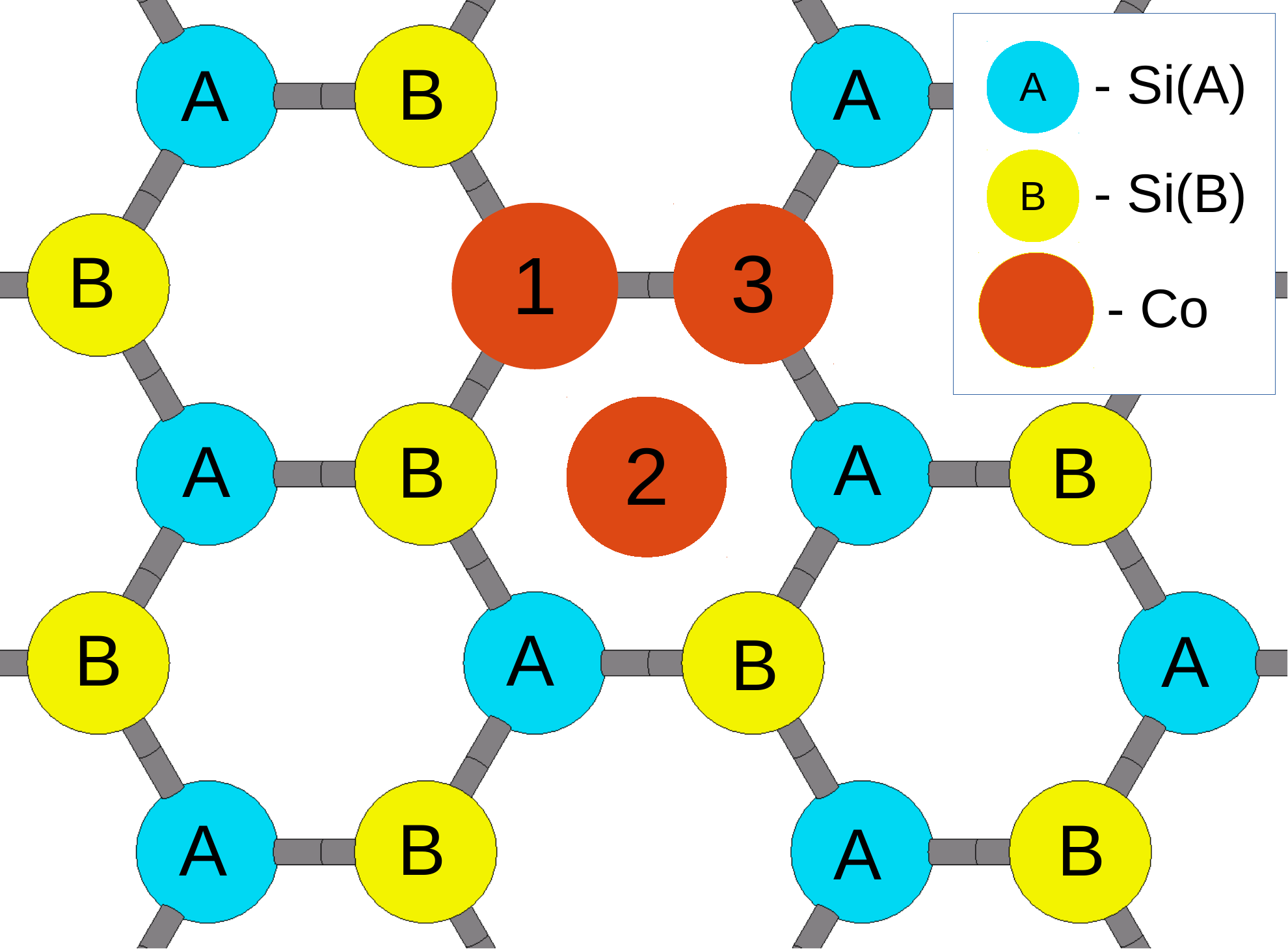}

\caption{Three possible high symmetry positions of Co adatom over the silicene
plane. The two silicene sublattices are marked as A and B with B being
the higher one. Note that because of the buckling the positions ``1''
and ``3'' are non-equivalent.}
\label{fig:Co_pos}
\end{figure}

The first principles calculations were performed using the generalized gradient approximation
(GGA) of density functional theory (DFT). The specific method applied
was full potential linearized augmented plane wave (FLAPW) \citep{singh:flapw2006}
as implemented in the WIEN2K package \citep{Blaha:wien2k}. The Perdew-Burke-Ernzerhof
parametrization \citep{Perdew_mz:prl96} of the exchange potential
was used in all the cases. The 2D Brilloiun zone (2D BZ) integration
was performed using the mesh densities corresponding to several hundreds
k-points (or more) in the single unit 2D BZ. The convergence criteria
for energy, charge per atom and forces were set to $\unit[10^{-4}]{Ry}$,
$\unit[10^{-3}]{e}$ and $\unit[2]{mRy/a.u.}$, respectively. In
all the calculations the silicene planes were separated by $\unit[14]{\mathring{A}}$
ensuring the lack of hopping between the neighboring planes.

In the first step the lattice constant and the sublattices'
displacement of bulk silicene were optimized. For the lowest energy
configuration we have found the in-plane lattice constant and the
vertical displacement to be equal to $a=\unit[3.86]{\mathring{A}}$ and
$\Delta=\unit[0.46]{\mathring{A}}$ \footnote{This corresponds to the
interatomic distance of $d_{Si-Si}=\unit[2.28]{\mathring{A}}$.},
respectively, in good agreement with the previous calculations
\citep{Cahangirov:prl2009,Liu:prl2011}.  When the spin-orbit
interactions were included in the calculations the small band gap of
$\Delta E\approx\unit[1.5]{meV}$ separating the tips of the Dirac
cones appeared in the band structure, also in good agreement with the
literature \citep{Liu:prl2011}.

Next we compared three possible high symmetry locations of the Co
adatom over the silicene. These are indicated in Fig.~\ref{fig:Co_pos}
and include the locations over two non-equivalent lattice sites (``1''
and ``3'') and also the position over the center of the hexagon
(``2''). The calculations were performed using 3x3 supercells with the
full relaxation of the atomic positions within the supercell.  It has
been found, in agreement with earlier works
\cite{Lin:prb12,Kaloni2014Jan}, that the order of the total energies is
as follows
\[
  E_{2}<E_{1}<E_{3}
\]
that is the central position ``2'' is the most favorable
energetically, followed by the position over the ``lower'' sublattice
A. The ground state is separated from the other two configurations by
$\Delta E_{\rm Tot}\approx\unit[0.17]{Ry}$.  These conclusions were
additionally confirmed by 4x4 calculations for ``1'' and ``2'' cases
with the same overall results.

\begin{figure}[b]
\includegraphics[width=0.98\columnwidth]{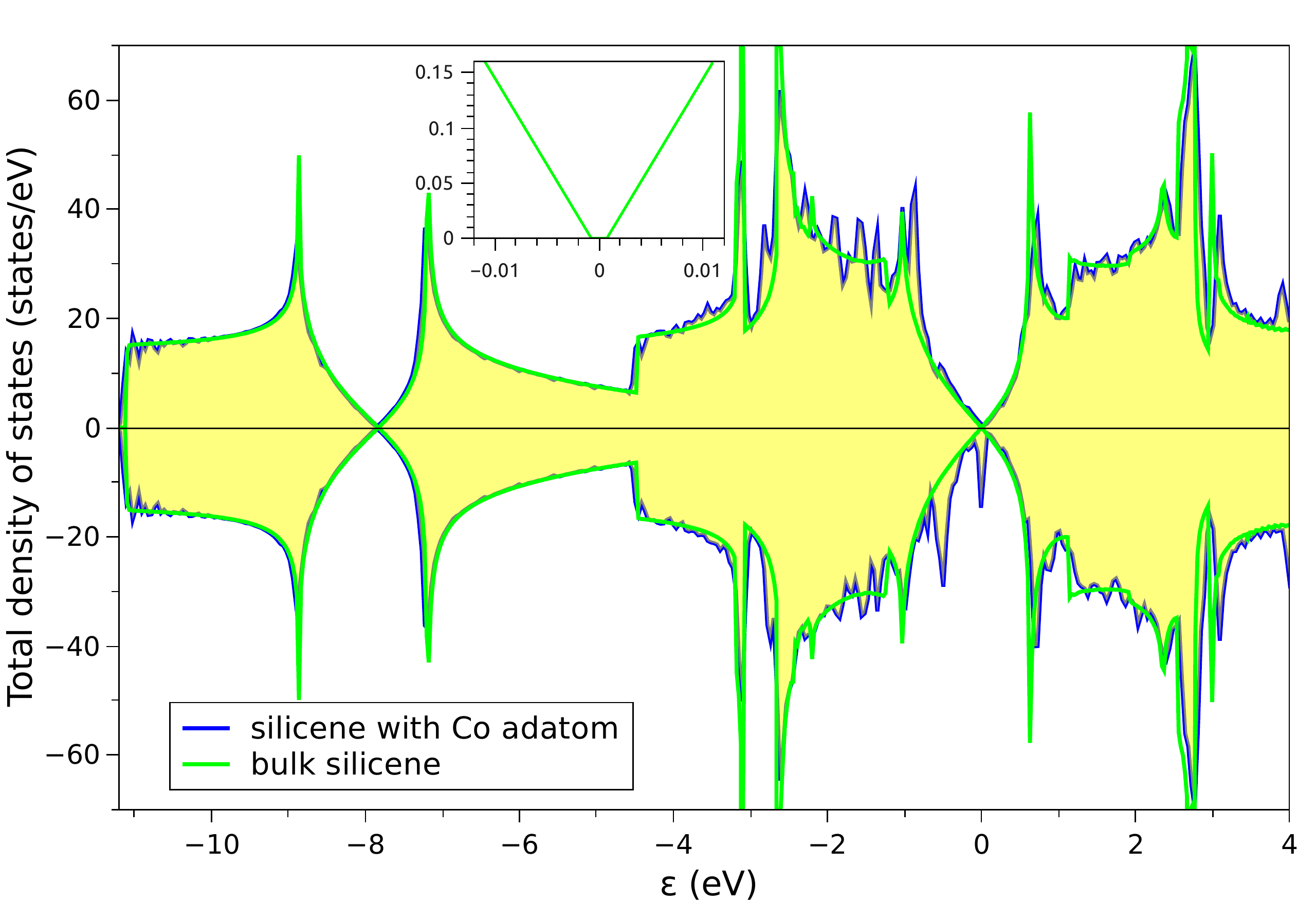}
\caption{The total density of states for a 7x7 supercell with
  centrally located Co adatom on silicene.  The green line corresponds
  to the density of states of pure silicene.  The Fermi energy
  corresponds to $\varepsilon=0$. The inset shows a small gap induced
  by the spin-orbit interaction in pure silicene.}
\label{fig:DOS_tot}
\end{figure}

In order to make sure that the size of the supercell used is
sufficiently large so that the limit of the single impurity is reached
we next studied the convergence of the magnetic moment and the local
density of states (LDOS) of the Co adatom against the size of the
supercell.  Because of the huge computational costs involved, the
structural relaxation was not performed during these
calculations. Instead we adopted the geometry obtained in 4x4
calculations for Co and its surroundings (up to
$R=\unit[8] {\mathring{A}}$ radius, corresponding to 4 coordination
zones) embedding so defined ``cluster'' into the bulk silicene. As the
deviations of Si atoms from bulk positions were found to be negligible
at this distance from Co, the procedure is reasonable. We have found
that both quantities stopped changing meaningfully when 7x7
(corresponding to the supercell lattice constant
$a_{sc}=\unit[27]{\mathring{A}}$) or larger supercells were used.  The
total magnetic moment, located predominantly on Co, equals
$\unit[1]{\mu_{B}}$.  We conclude therefore that the impurity can be
effectively treated as a spin one-half in the Anderson model.

\begin{figure}[t]
\includegraphics[width=0.98\columnwidth]{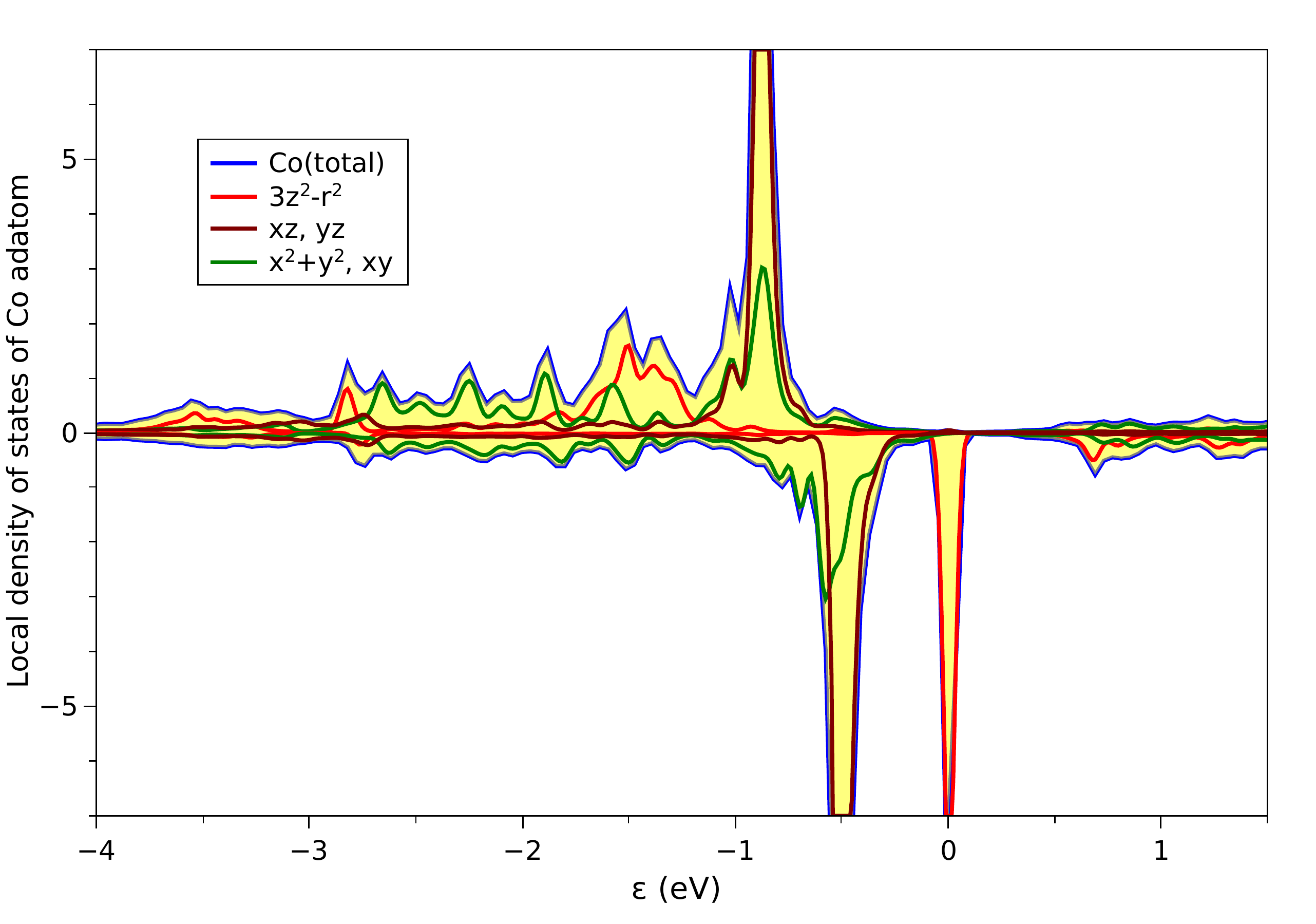}
\caption{The local density of states of Co adatom. Both total and
  partial, orbital resolved contributions are indicated.}
\label{fig:LDOS_Co}
\end{figure}

In the final step we have calculated global and local DOS using 7x7
supercell and including the spin-orbit coupling. The 2D BZ integration
was performed at this point using suitably dense mesh corresponding to
$10^{4}$ k-points in the original 1x1 2D BZ. The results for the total
DOS of the structure, together with the DOS of pure silicene,
are presented in Fig.\,\ref{fig:DOS_tot}.  In the
vicinity of the adatom the hybridization effect comes into play, and
it drastically rebuilds the LDOS. In consequence, the DOS of the
system with an adatom follows the same general outline as the DOS of
pure silicene but with additional modulations visible in
Fig.~\ref{fig:DOS_tot}. The most notable of these is perhaps the sharp
peak located exactly at the Fermi energy in the minority spin channel.
Analysis of the orbital contribution to the local density of states of
cobalt shown in \fig{fig:LDOS_Co} indicates that the peak at the Fermi
energy in the total DOS of the structure originates mainly from the
$d_{3z^{2}-r^2}$ orbitals of the adatom.

\section{Numerical renormalization group calculations}

\subsection{Effective model}

Based on the first-principle results we can now establish an effective
Hamiltonian for the Co adatom on silicene.  The calculated magnetic
moment of adatom justifies usage of a spin one-half single impurity
Anderson model \cite{Anderson1961Oct}. In order to find the parameters
of the effective single-orbital model we follow the procedure
described in Ref.~[\onlinecite{Ujsaghy2000Sep}].  From the calculated
occupancy of Co orbitals, $n_d\approx 7.8$ (in agreement with
Ref. [\onlinecite{Kaloni2014Jan}]), we conclude that charge fluctuations
can occur between the states with $7$, $8$ and $9$ electrons.
Starting with the Anderson model with full five-fold degeneracy of the
d-shell electrons, we note that in the mean field approximation
the energies of the respective charge states enumerated by $j$ can be
expressed as, $E(j) = j \epsilon +Uj(j-1) /2 $, where $\epsilon$ is
an adatom on-site energy and $U$ denotes the Coulomb correlations.
Following Ref.~[\onlinecite{Kaloni2014Jan}], the latter parameter will
be set equal to $U=4$ eV.
From the minimum energy condition with respect to $j$ we can estimate the
on-site energy $\epsilon = (1/2-n_d)U$.  Focusing on consecutive energies of
the states with $7$, $8$ and $9$ electrons (with respect to the lowest
energy), we can find from $\tilde\epsilon_d = E(8)-E(7)$ and
$2 \tilde\epsilon_d + \tilde U = E(9)-E(7)$, the parameters for
the effective impurity Hamiltonian, $\tilde \e_d=-1.2$ eV,
$\tilde U=U$, which has the following form
\be
H = H_{\rm band} + H_{\rm imp} + H_{\rm tun}.
\ee
Here, $H_{\rm band} = \sum_{\sigma} \int d\e \,\e \,c_{\sigma}^\dagger(\e) c_{\sigma}(\e)$
describes the electrons in silicene with the corresponding density of states, see \fig{fig:DOS_tot}.
The second term models adatom and is given by
$H_{\rm imp} = \sum_\sigma \tilde \e_d d_\sigma^\dag d_\sigma
+ \tilde U d_\up^\dag d_\up d_\down^\dag d_\down + g\mu_{\rm B} S_z B$,
where $\tilde \e_d$ is the energy of an electron occupying the impurity
and $\tilde U$ denotes the Coulomb correlation energy.
The last term accounts for the Zeeman splitting, with $B$ being the external magnetic field
and $S_z$ denoting the spin of the adatom.
The operator $d^\dag_\sigma$ creates a spin-$\sigma$ electron on the adatom,
and $c^\dag_{\sigma}(\e)$ is the corresponding creation operator for spin-$\sigma$
electrons in silicene.
Finally, the coupling between the substrate and the adatom is modeled
by the tunneling Hamiltonian
$H_{\rm tun} = \sum_{\sigma} \int d\e V \sqrt{\rho (\e)}
[c^\dag_\sigma(\e) d_\sigma + d^\dag_\sigma c_\sigma(\e)] $,
where $V$ denotes the tunnel matrix elements assumed to be equal to $V=0.65$ eV
and $\rho(\e)$ is the density of states of bulk silicene, cf. \fig{fig:DOS_tot}.
The parameter $V$ has been evaluated by
Harrison's scaling method \cite{Harrison} for
the average Si-Co bond length $r=2.45$ \r{A}.

Since we are interested in nonperturbative effects resulting
from the hybridization of Co adatom and silicene,
to get the most accurate information about system's spectral properties
we employ the numerical renormalization group (NRG) method
\cite{Wilson_Rev.Mod.Phys.47/1975,Legeza_DMNRGmanual,Bulla_RMP.80/395}.
In NRG, the band is first discretized in a logarithmic way
with a discretization parameter $\Lambda$. Then,
such discretized Hamiltonian is tridiagonalized numerically
and transformed to a tight-binding Hamiltonian of the following form
\bea
H_{\rm NRG} = H_{\rm imp} + \sum_\s V (d_\sigma^\dag f_{0\sigma} +  f_{0\sigma}^\dag d_\sigma)\nonumber\\
+ \sum_{n=0}^\infty \sum_\s \left[ \epsilon_n f_{n\sigma}^\dag f_{n\sigma}
+  t_n (f_{n\sigma}^\dag f_{n+1\sigma} + f_{n+1\sigma}^\dag f_{n\sigma} ) \right] \!,
\eea
where $f_{n\sigma}^\dag$ is the creation operator of an electron
with spin $\s$ on the $n$th site of the chain,
$t_n$ denotes the hopping integral and $\epsilon_n$ is the on-site energy.
In NRG calculations we assumed $\Lambda=1.7$ and kept at least $1500$
states at each iteration. To obtain the most accurate results for the spectral functions,
we also optimized the broadening parameter appropriately
\cite{Zitko2009Feb}.

\subsection{Discussion of numerical results}

\begin{figure}[t]
\includegraphics[width=0.95\columnwidth]{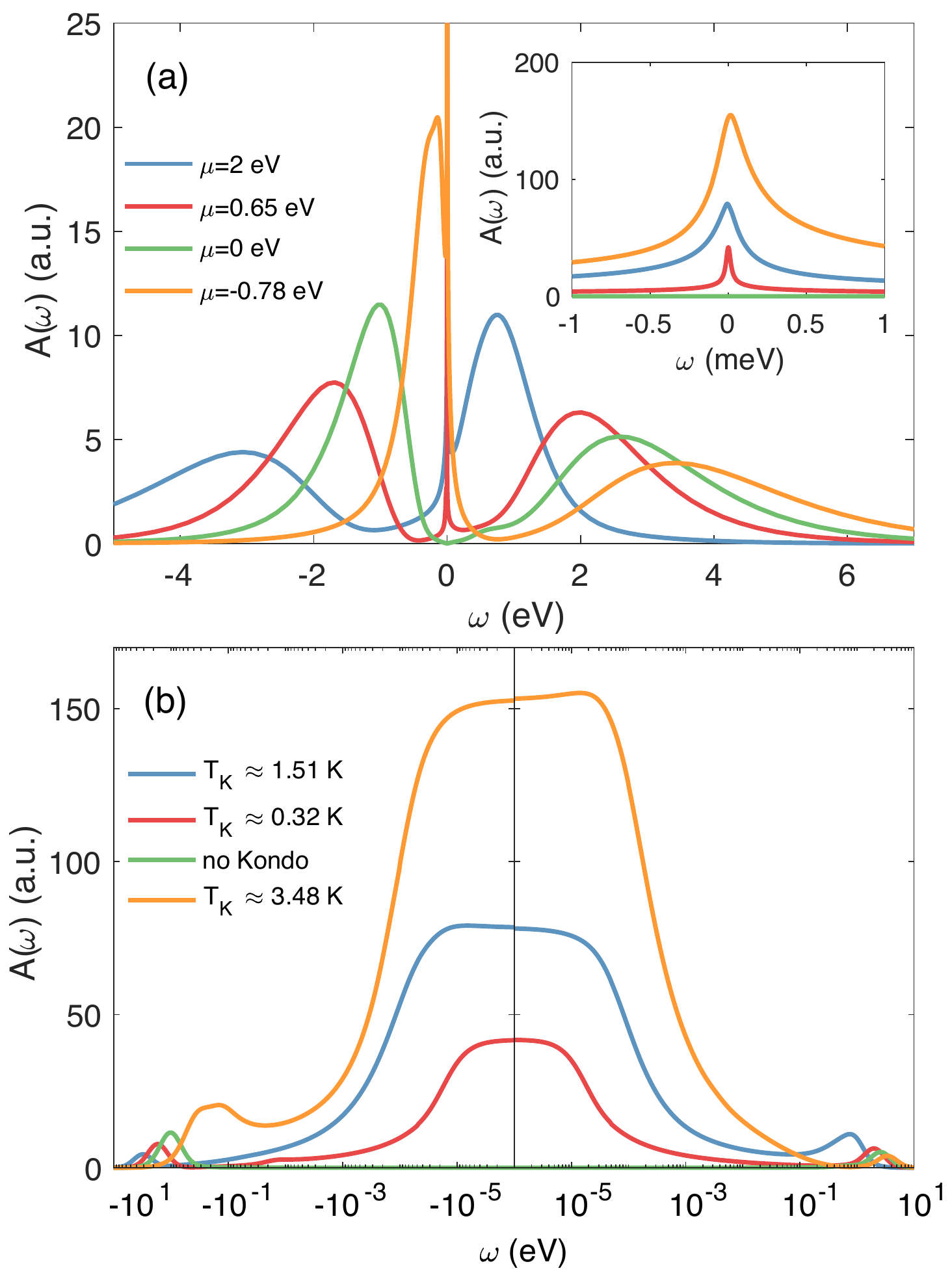}
\caption{
The zero-temperature total spectral function
of the Co adatom calculated by NRG for 
a few values of chemical potential $\mu$, $E_F = \mu$,
plotted on (a) linear and (b) logarithmic scale.
The inset in (a) presents the zoom onto the low-energy
behavior where the Kondo effect can emerge.
The associated Kondo temperatures, obtained
from the halfwidth at half maximum
of the resonance in the spectral function,
are indicated in the legend of panel (b).
The parameters are: $\tilde U=4$ eV, $\tilde \e_d = -1.2$ eV,
$V=0.65$ eV, and $B=0$. We also set $\hbar \equiv 1$.}
\label{fig:Amu}
\end{figure}

We now focus on the behavior of the spectral function of Co adatom,
$A_\sigma(\omega) = -{\rm Im} G^R_\sigma(\omega)/\pi$, where
$G^R_\sigma(\omega)$ is the Fourier-transform of the corresponding retarded Green's function,
$G^R_\sigma(t) = -i\Theta(t)\langle \{  d_\sigma(t), d_\sigma^\dag(0) \} \rangle$.
The total spectral function, $A(\omega) = A_\uparrow(\omega) + A_\downarrow(\omega)$,
corresponds to the local density of states
of adatom, which can be experimentally examined with a weakly coupled
probe, such as a tip of a scanning tunneling microscope.

\subsubsection{Spectral properties and the Kondo effect}

Because experimentally the position of the Fermi level can be adjusted by gating,
in \fig{fig:Amu} we present the energy dependence of $A(\omega)$ 
calculated for different values of the chemical potential $\mu$, $E_F = \mu$.
Let us first focus on the case of no gating, $\mu=0$.
One can see that the total spectral function exhibits
Hubbard resonances for $\omega = \tilde \e_d$ and $\omega = \tilde \e_d + \tilde U$.
For typical spin one-half quantum impurity models, at low energies, the
Kondo physics plays an important role
\cite{Kondo_Prog.Theor.Phys32/1964}. In the Kondo effect the
conduction electrons screen the impurity's spin resulting in an
additional resonance at the Fermi energy in the local density of
states, the halfwidth of which is related to the Kondo temperature $T_K$
\cite{Hewson_book}. In the case considered here, however, due to the
depletion of states at the Fermi energy the screening of the adatom
spin is not possible and, consequently, the Kondo peak is not present,
see \fig{fig:Amu} for $\mu=0$.

One can consider if it is possible to reinstate
the Kondo resonance by changing the chemical potential via gating.
First of all, it can be seen that, quite naturally, by tuning $\mu$
the position of the Hubbard resonances changes accordingly, see  \fig{fig:Amu}.
Moreover, by adjusting the Fermi energy,
one can also considerably affect the low-energy behavior of the system.
In fact, for values of $\mu$ selected in \fig{fig:Amu},
a pronounced Kondo resonance develops.
This can be clearly seen in \fig{fig:Amu}(b),
which shows the spectral function plotted on logarithmic scale,
as well as in the inset of \fig{fig:Amu}(a),
which presents a zoom into the low energy behavior of $A(\w)$.
For the considered values of gating, there is sufficient number
of states at the Fermi energy to screen the adatom's spin.
One can then observe the Kondo effect
with relatively large Kondo temperature, of the order of up to a few Kelvins,
as estimated from the halfwidth at half maximum of the Kondo peak,
see \fig{fig:Amu}(b).

\begin{figure}[t]
\includegraphics[width=1\columnwidth]{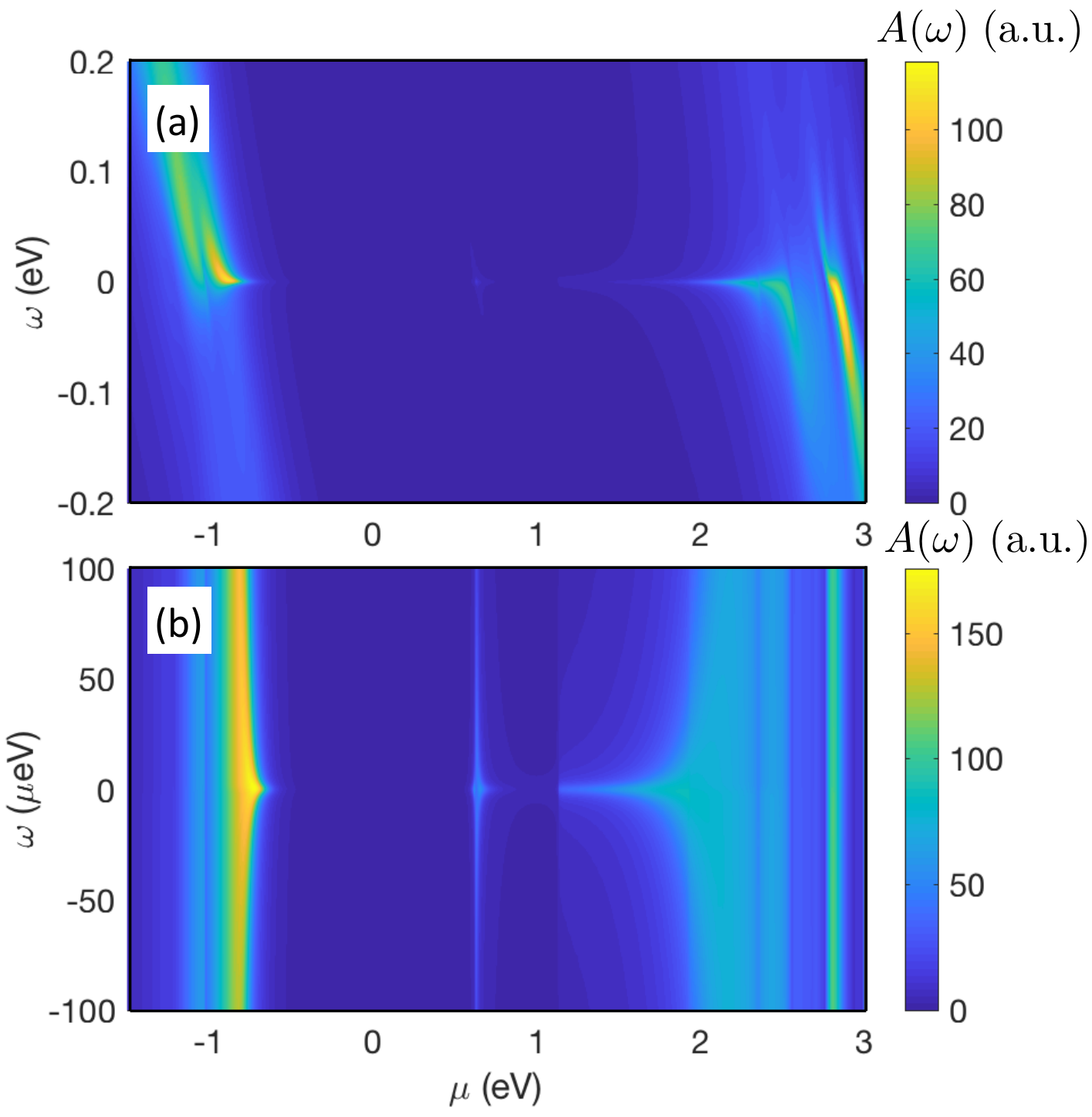}
\caption{
The total spectral function plotted versus energy $\omega$
and the chemical potential $\mu$,
$E_F = \mu$, calculated for parameters the same as in \fig{fig:Amu}.
The bottom panel (b) shows the zoom of the low-energy behavior of $A(\omega)$.}
\label{fig:Amu2D}
\end{figure}

\begin{figure}[t]
\includegraphics[width=0.95\columnwidth]{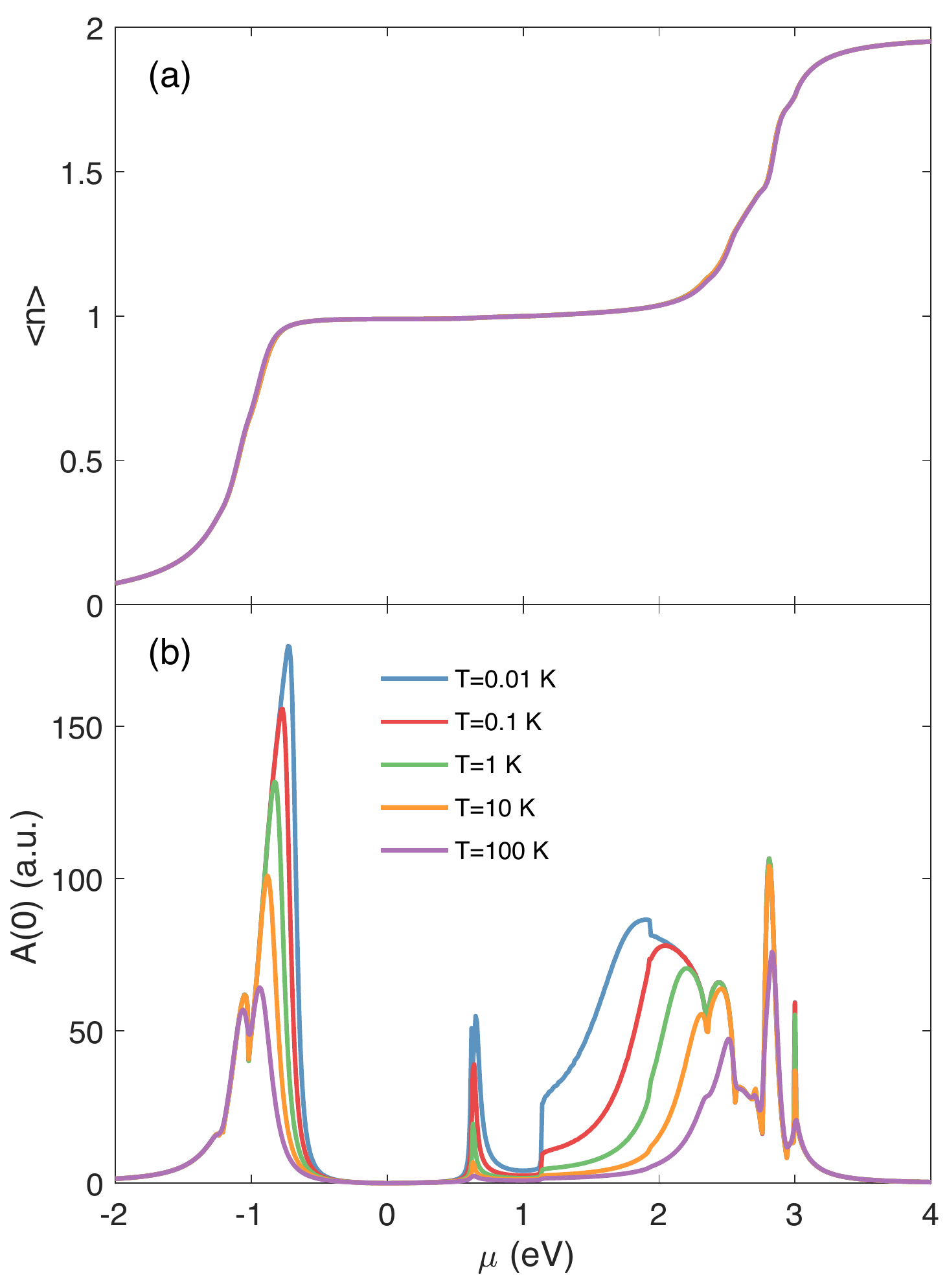}
\caption{
(a) The occupation of the orbital level $\langle n\rangle$
and (b) the spectral function at $\w=0$ $A(0)$ plotted as a function of chemical potential $\mu$
and calculated for different temperatures, as indicated.
The other parameters are the same as in \fig{fig:Amu}.
Note that the occupation hardly depends on temperature,
since the energy scales of the adatom are still much larger
than the considered temperatures.
}
\label{fig:A0T}
\end{figure}

To gain a deeper understanding of the effect of gating on the local density of states,
we also analyze the behavior of the spectral function
by continuously tuning the chemical potential. This is presented
in \fig{fig:Amu2D}, which shows the energy and chemical 
potential dependence of the local density of states, with the bottom panel zooming
into the low-energy behavior of $A(\omega)$.
First, we note that for $\mu \approx -1.2$ eV and $\mu \approx 2.8$ eV,
the spectral function exhibits maxima related with resonant tunneling,
since then the empty and doubly occupied states of the adatom cross the Fermi energy
(the adatom enters the mixed valence regime).
In the range of chemical potential between these two values,
the impurity effectively hosts a spin one-half,
see \fig{fig:A0T}(a) presenting the chemical potential dependence of 
the occupation $\langle n \rangle$ of the orbital level, 
where $n = \sum_\sigma d^\dag_\sigma d_\sigma$.
Consequently, one can expect that the Kondo effect will emerge
if the number of states is sufficient to screen the impurity's spin.
The occurrence of the Kondo resonance
is thus strongly dependent on the hybridization,
which changes with $\mu$.
As a matter of fact, one can indeed clearly identify regions 
where the resonance at the Fermi energy develops.
These regions are better visible in \fig{fig:Amu2D}(b), which
shows the zoom into the low-energy behavior of the spectral function.
From the inspection of this figure
one can see that for values of chemical potential
ranging from $\mu \approx 1.2$ eV to  $\mu \approx 2.5$ eV,
the Kondo resonance develops with Kondo temperature strongly dependent on $\mu$.
The Kondo temperature becomes enhanced for $\mu \approx 2$ eV,
where $T_K\approx 1.5$ K.
Moreover, a relatively large Kondo temperature
can be also found for $\mu = -0.78$ eV, where $T_K\approx 3.48$ K, cf. \fig{fig:Amu}(b).

We would like to recall that the Kondo temperature
depends in an exponential fashion on the ratio of hybridization
between the orbital level and the host
and the Coulomb correlations \cite{Hewson_book}.
Thus, from theoretical point of view, if the host's density of states
at the Fermi energy is finite, the Kondo peak should emerge even if DOS is very low.
However, the corresponding Kondo temperature would be then extremely small
and, thus, the Kondo peak would be completely undetectable experimentally.
In \fig{fig:A0T}(b) we present the chemical potential dependence
of the spectral function at $\omega=0$ for several,
experimentally relevant, values of temperature.
By looking at the orbital level occupation, \fig{fig:A0T}(a),
one can recognize enhanced spectral function due to resonant
tunneling in the mixed valence regime
and due to the Kondo effect in the local moment regime.
Moreover, one can also inspect how quickly the Kondo correlations
become smeared out by thermal fluctuations,
which is especially visible in the range of $\mu$
from $\mu \approx 1.2$ eV to $\mu \approx 2.5$ eV.

\subsubsection{Effect of external magnetic field}

\begin{figure}[t]
\includegraphics[width=0.95\columnwidth]{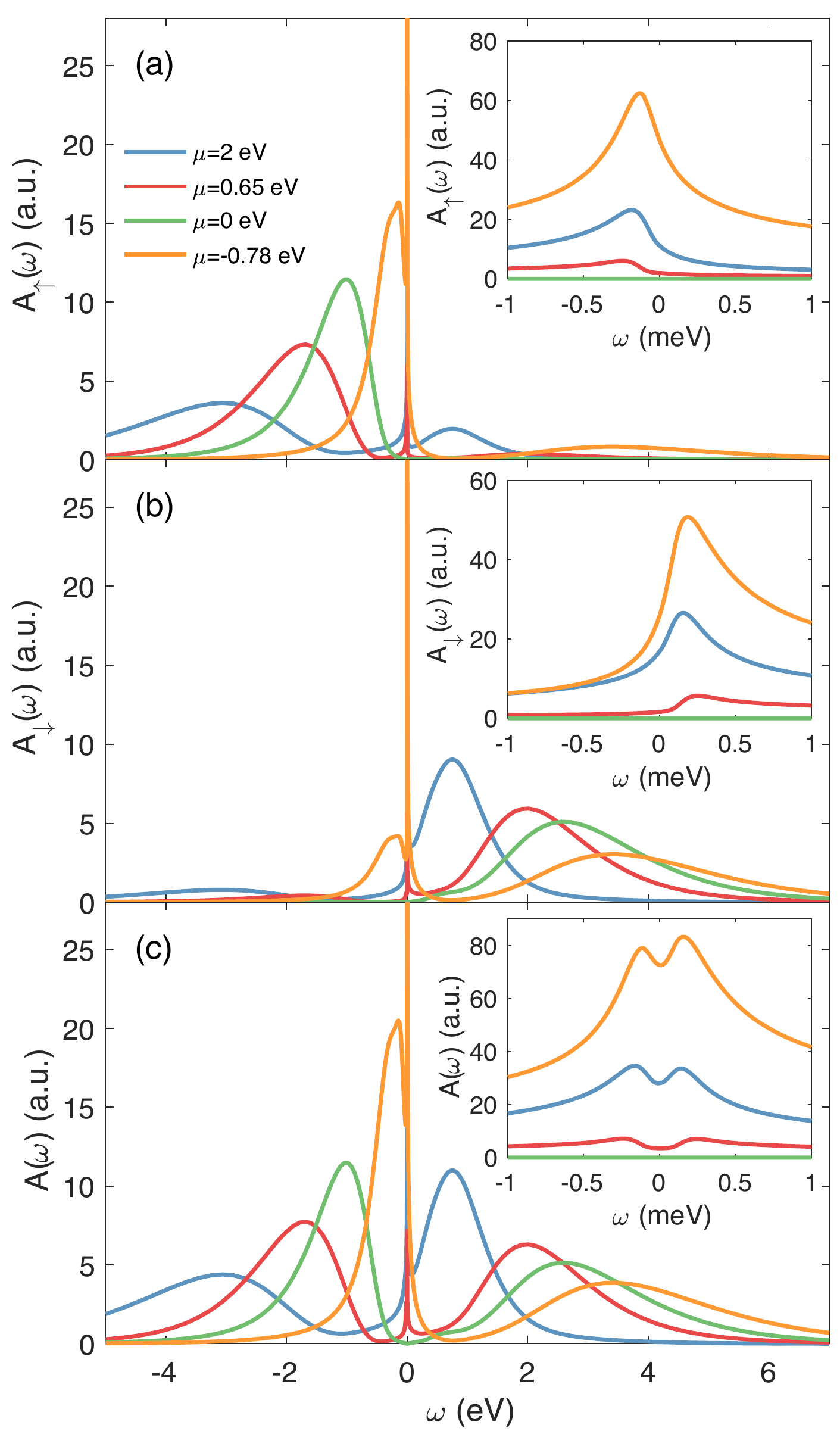}
\caption{
The zero-temperature (a) spin-up, (b) spin-down
and (c) total spectral function
for different values of chemical potential,
as indicated in the figure,
calculated in the presence of external magnetic field $B = 2$ T.
The insets present the zoom into the low-energy
behavior of the spectral function.
Parameters are the same as in \fig{fig:Amu}
and we assumed the $g$-factor, $g=2$.}
\label{fig:ABz}
\end{figure}

\begin{figure}[t]
\includegraphics[width=1\columnwidth]{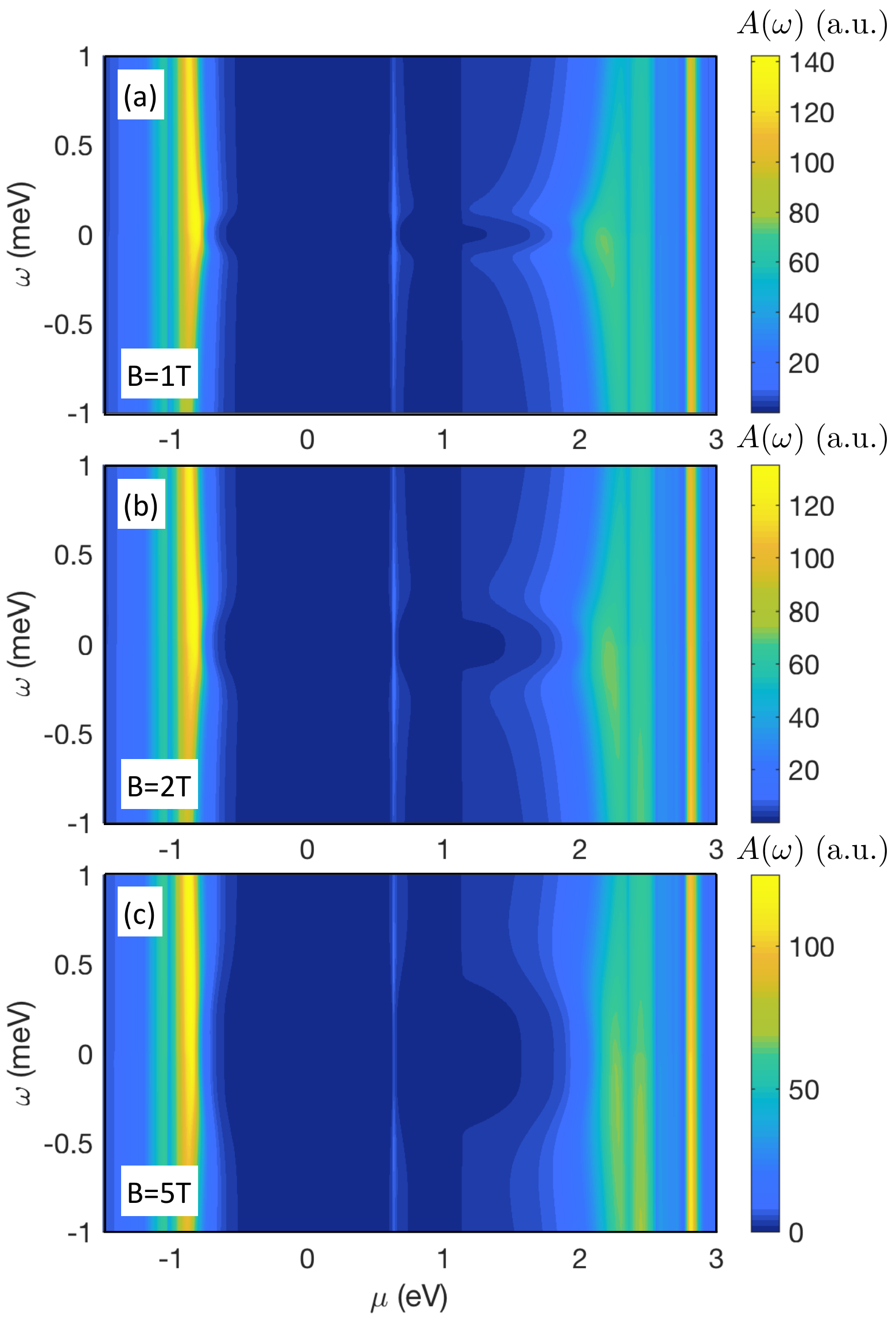}
\caption{
The energy and chemical potential dependence of the total spectral function
for different values of external magnetic field:
(a) $B=1$ T, (b) $B=2$ T and (c) $B=5$ T.
The other parameters are the same as in \fig{fig:Amu}.}
\label{fig:ABz2D}
\end{figure}

\begin{figure}[t]
\includegraphics[width=1\columnwidth]{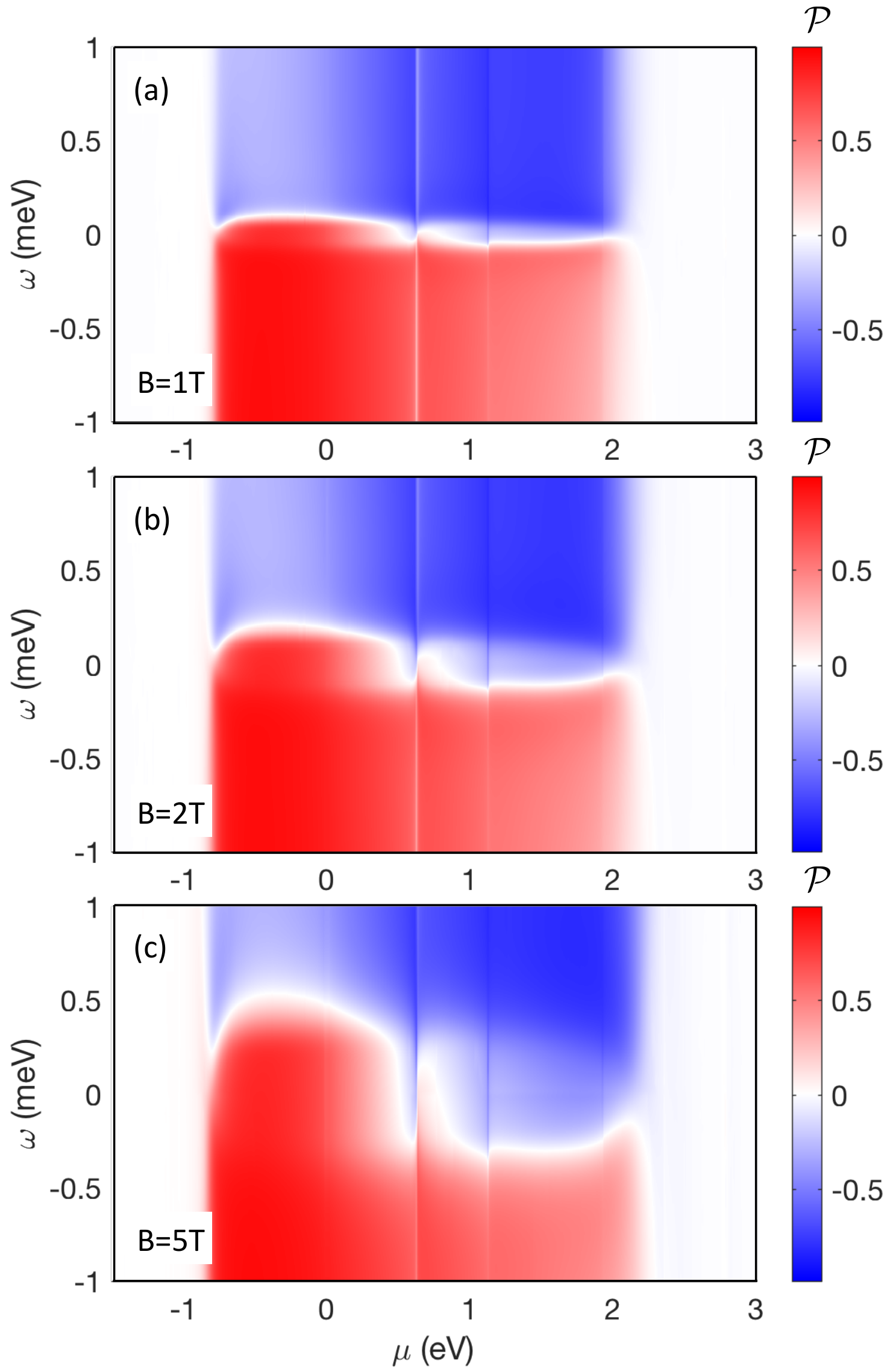}
\caption{
The energy and chemical potential dependence of
the spin polarization of the spectral function,
$\mathcal{P} = [A_\uparrow(\omega) - A_\downarrow(\omega)]/A(\omega)$,
calculated for different values of external magnetic field:
(a) $B=1$ T, (b) $B=2$ T and (c) $B=5$ T.
The other parameters are the same as in \fig{fig:Amu}.}
\label{fig:P2D}
\end{figure}

Let us now consider how external magnetic field affects
the behavior of the local density of states of Co adatom.
The energy dependence of the spin-resolved 
and total spectral function calculated in the presence
of external magnetic field $B=2$ T is depicted
in \fig{fig:ABz}, where the insets show the zoom into
the low-energy behavior of $A(\w)$.
First of all, one can see that magnetic field strongly affects the behavior of the spectral function.
An important observation is a strong spin polarization of Hubbard resonances:
$A_\uparrow$ ($A_\downarrow$) becomes suppressed for $\w>0$ ($\w<0$).
Moreover, magnetic field also has a strong effect on
the low-energy behavior of $A(\w)$:
the Kondo resonance becomes split if the Zeeman energy $E_Z$ becomes larger than 
$k_B T_K$, $E_Z = g\mu_{\rm B} B\gtrsim k_B T_K$.
The spin polarization of the spectral function
can be clearly seen in Figs. \ref{fig:ABz}(a) and (b),
while the splitting of the Kondo peak is nicely visible in the inset of \fig{fig:ABz}(c).
We note that the largest suppression of the Kondo peak
occurs for $\mu = 0.65$ eV. For this value of gating
we estimated $T_K \approx 0.32$ K, which is smaller
than for the case of $\mu = 2$ eV and $\mu = -0.78$ eV.
Consequently, larger suppression of the Kondo resonance
is observed for smaller $T_K$, since the condition $E_Z \gtrsim k_B T_K$ is then better satisfied.
We also notice that the effect of splitting and suppression of the Kondo peak
in the presence of magnetic field is in fact similar to the effect of an
exchange-field splitting of orbital level caused by the presence
of ferromagnetic correlations
\cite{Martinek2003Sep,Hauptmann2008May,Gaass2011Oct,Weymann2011Mar,Csonka2012May}.
Thus, if due to proximity effect with magnetic substrate,
the density of states of silicene becomes 
spin-polarized, the Kondo effect may be also split and suppressed
even in the absence of magnetic field.

It is also interesting to analyze the energy and chemical potential dependence of 
the local density of states for several values of external magnetic field.
This is presented in \fig{fig:ABz2D}, which shows $A(\w)$ for $B=1, 2,5$ T.
In this figure we focus on the low-energy regime, where Kondo effect can emerge,
and the interplay between the Zeeman splitting and Kondo correlations is most revealed.
One can see that, depending on the value of external magnetic field
and gating, the Kondo resonance can become split and suppressed.
This is especially visible for $\mu = 1.5$ eV, where with increasing $B$,
one shifts the position of the split Kondo peaks.
Moreover, there are such values of $\mu$, especially
those close to the mixed valence regime,
where the magnetic field is not strong enough to suppress the Kondo resonance,
see \fig{fig:ABz2D}.

The effect of external magnetic field can be better revealed when one considers
the spin polarization of the spectral function, which is defined as,
$\mathcal{P} = [A_\uparrow(\omega) - A_\downarrow(\omega)]/A(\omega)$.
The dependence of the spin polarization on energy $\omega$
and chemical potential $\mu$ for a few values of magnetic field is shown in \fig{fig:P2D}.
This figure is generated for the same values of $B$
as those considered in \fig{fig:ABz2D},
again focusing on the low-energy behavior.
When the impurity is either empty or doubly occupied,
the spin polarization is suppressed and approaches zero.
Its behavior, however, becomes completely changed in the local moment regime,
see \fig{fig:P2D}.
The first observation is that $\mathcal{P}$ can change sign
around $\omega = 0$ and such sign flip occurs in the regime
where the impurity is occupied by a single electron.
This effect is associated with the spin-splitting of the adatom orbital
due to the Zeeman field.
One can notice that for higher energies the spectral function becomes fully spin-polarized,
$\mathcal{P}\approx 1$ for $\w<0$ and $\mathcal{P}\approx -1$ for $\w>0$.
Interestingly, it can be also seen that the interplay of finite Zeeman splitting
and the density of states of silicene, can result in 
a sign change of the spin polarization at $\w=0$ around $\mu=0.5$ eV,
which is most visible for $B=5$ T, see \fig{fig:P2D}(c).
At low energies and for $\mu\lesssim 0.5$ eV ($\mu\gtrsim 0.5$ eV),
$\mathcal{P}$ becomes positive (negative).

\begin{figure}[t]
\includegraphics[width=1\columnwidth]{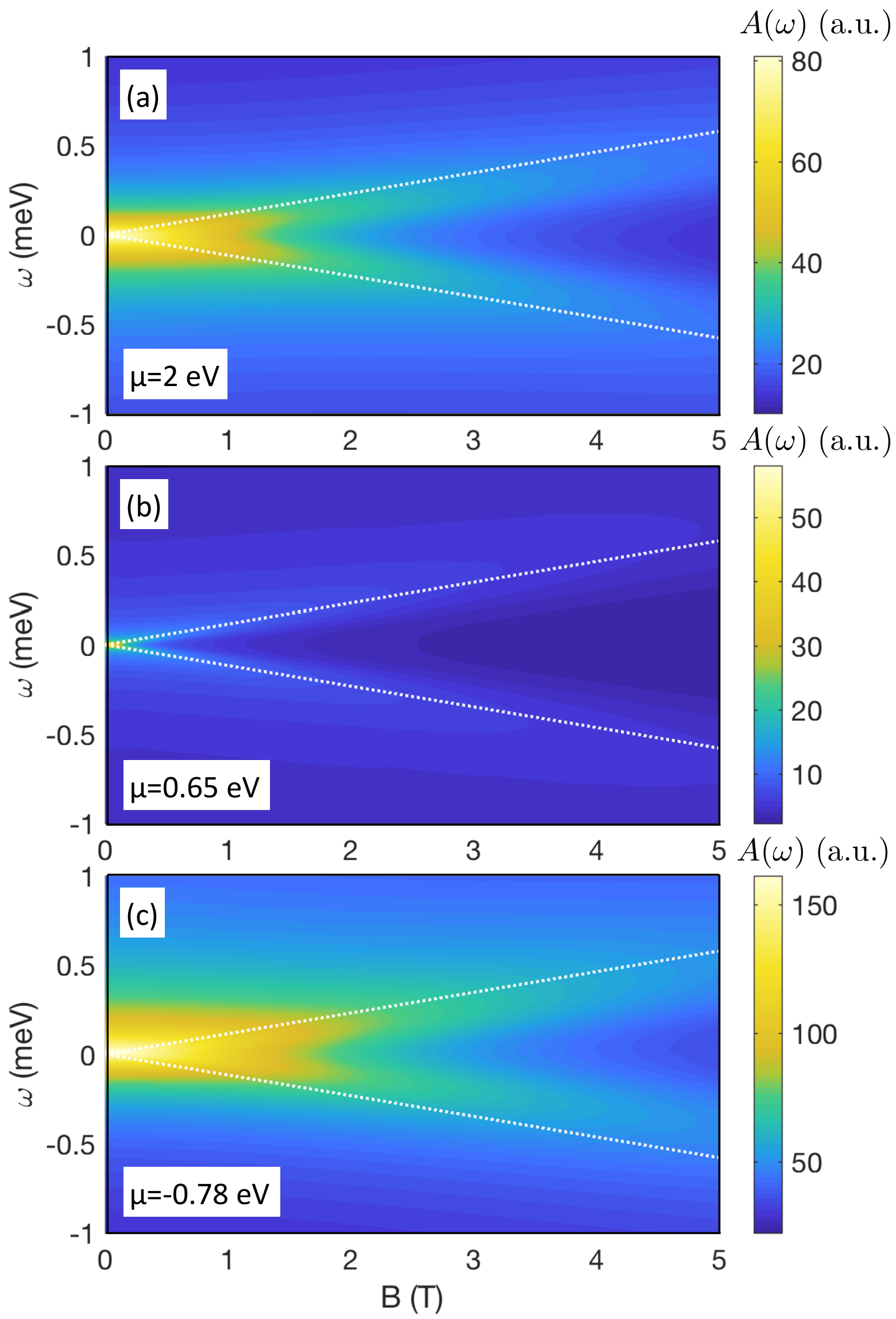}
\caption{
The spectral function plotted as a function of energy and magnetic field
for selected values of chemical potential for which pronounced Kondo peak
develops in the absence of magnetic field:
(a) $\mu=2$ eV, (b) $\mu=0.65$ eV and (c) $\mu=-0.78$ eV.
The dotted lines mark the Zeeman energy $E_Z = \pm g\mu_B B$.
The other parameters are the same as in \fig{fig:Amu}.}
\label{fig:Asplit}
\end{figure}

Finally, in \fig{fig:Asplit} we study the evolution of the splitting of the Kondo peak
with external magnetic field. This figure is calculated for selected
values of the chemical potential, the same as considered in \fig{fig:ABz}.
It can be seen that the Kondo peak becomes suppressed when magnetic field
is so strong that the condition $E_Z \gtrsim k_B T_K$ is fulfilled.
Thus, the suppression occurs
first for the case of $\mu=0.65$ eV, while larger field
is needed to destroy the Kondo resonance for the other two cases,
see \fig{fig:Asplit}.
When magnetic field is large enough such that $E_Z \gtrsim k_B T_K$,
$A(0)$ becomes suppressed and the spectral function
shows only side peaks, which occur exactly at the spin-flip
excitation energy $\omega = \pm E_Z$.
The position of those side peaks depends thus linearly on magnetic field,
which can be nicely seen in \fig{fig:Asplit},
where the dotted lines mark the Zeeman energy $E_Z$.

\section{Conclusions}

We have theoretically considered the spectroscopic properties
and the Kondo effect of Co adatoms on silicene.
By using the first principle calculations, we have determined the total
density of states of Co-silicene system and estimated the orbital level occupancy
together with the magnetic moment of Co. Our DFT results allowed us to formulate
an effective low-energy Hamiltonian for spin one-half impurity,
which was further used to analyze the spectral properties of Co adatoms.
This analysis was performed by employing the numerical renormalization
group method with a non-constant density of states.
We focused on the behavior of the local density of states (spectral function),
which can be probed experimentally by using the scanning tunneling spectroscopy.
The analysis involved the effects of external magnetic field
and gating of silicene. We showed that by appropriately tuning
the parameters one can obtain clear signatures of the Kondo effect.
We also analyzed the evolution and splitting of the Kondo resonance
with an external magnetic field.
Finally, we studied the spin polarization of the spectral function
in the presence of magnetic field,
whose magnitude and sign were found to
greatly depend on the position of chemical potential.


\begin{acknowledgments}
I.W. acknowledges discussions with C. P. Moca.
This work is supported by Polish National Science Centre from funds
awarded through the decision No. DEC-2013/10/M/ST3/00488.
\end{acknowledgments}



%

\end{document}